\documentclass[sigconf, 10pt]{acmart}

\usepackage{graphicx} 
\usepackage{kotex}   
\usepackage{comment}
\usepackage{xspace}
\usepackage{bm}
\usepackage[normalem]{ulem}
\usepackage[linesnumbered,ruled,vlined]{algorithm2e}
\usepackage{multirow}
\usepackage{colortbl}
\usepackage{multicol}
\usepackage{graphics}
\usepackage{titlesec}
\usepackage{simplebnf} 

\titlespacing*{\subsubsection}{0pt}{0.1\baselineskip}{0.2\baselineskip}
\titlespacing*{\subsection}{0pt}{0.2\baselineskip}{0.2\baselineskip}
\titlespacing*{\section}{0pt}{0.4\baselineskip}{0.2\baselineskip}

\usepackage{color}
\definecolor{gray}{rgb}{0.4,0.4,0.4}
\definecolor{darkblue}{rgb}{0.0,0.0,0.6}
\definecolor{cyan}{rgb}{0.0,0.6,0.6}
\definecolor{maroon}{rgb}{0.5,0,0}
\definecolor
{darkgreen}{rgb}{0,0.5,0}
\definecolor{lightgray}{rgb}{0.8,0.8,0.8}

\usepackage{xcolor}
\definecolor{codegreen}{rgb}{0,0.6,0}
\definecolor{codegray}{rgb}{0.5,0.5,0.5}
\definecolor{codepurple}{rgb}{0.58,0,0.82}
\definecolor{backcolour}{rgb}{0.95,0.95,0.92}

\usepackage{listings}
\lstdefinestyle{mystyle}{
    frame=tb,
    showstringspaces=false,
    columns=fullflexible,
    commentstyle=\color{codegreen},
    keywordstyle=\color{magenta},
    numberstyle=\tiny\color{codegray},
    numbers=none,
    stringstyle=\color{codepurple},
    basicstyle=\footnotesize,
    breakatwhitespace=false,
    breaklines=true,
    captionpos=b,
    keepspaces=true,
    showspaces=false,
    showstringspaces=false,
    showtabs=false,
    tabsize=2
}

\lstdefinelanguage{XML}{
    frame=tb,
    aboveskip=1mm,
    belowskip=1mm,
    showstringspaces=false,
    basicstyle=\footnotesize,
    morestring=[s]{"}{"},
    morecomment=[s]{?}{?},
    morecomment=[s]{!--}{--},
    commentstyle=\color{gray}\upshape,
    moredelim=[s][\color{black}]{>}{<},
    stringstyle=\color{black},
    morekeywords={index, id, text, description, UI_index},
    identifierstyle=\color{darkblue},
    keywordstyle=\color{cyan},
    tabsize=1,
    keepspaces=true,
    showstringspaces=false,
    showtabs=false,
    moredelim=**[is][\bfseries]{@}{@},
}
\definecolor{delim}{RGB}{20,105,176}
\definecolor{numb}{RGB}{106, 109, 32}
\definecolor{string}{rgb}{0.64,0.08,0.08}
\lstdefinelanguage{json}{
    frame=tb,
    aboveskip=1mm,
    belowskip=1mm,
    showstringspaces=false,
    basicstyle=\footnotesize,
    rulecolor=\color{string},
    breaklines=true,
    commentstyle=\color{gray}\upshape,
    breakatwhitespace=true,
    morestring=[s]{"}{"},
    stringstyle=\color{black},
    keywordstyle=\color{string},
    morekeywords={name, desc, params, UI_index, UI_attrib, id, requires},
    moredelim=**[is][\bfseries]{@}{@},
    literate=
     *{0}{{{\color{numb}0}}}{1}
      {1}{{{\color{numb}1}}}{1}
      {2}{{{\color{numb}2}}}{1}
      {3}{{{\color{numb}3}}}{1}
      {4}{{{\color{numb}4}}}{1}
      {5}{{{\color{numb}5}}}{1}
      {6}{{{\color{numb}6}}}{1}
      {7}{{{\color{numb}7}}}{1}
      {8}{{{\color{numb}8}}}{1}
      {9}{{{\color{numb}9}}}{1}
      {\{}{{{\color{delim}{\{}}}}{1}
      {\}}{{{\color{delim}{\}}}}}{1}
      {[}{{{\color{delim}{[}}}}{1}
      {]}{{{\color{delim}{]}}}}{1},
}

\newcommand{\syslong}{\textsf{VeriSafe Agent}\xspace}
\newcommand{\sys}{\textsf{VSA}\xspace}
\newcommand{\syswarm}{\textsf{VSA-Warm}\xspace}
\newcommand{\syscold}{\textsf{VSA-Cold}\xspace}

\newcommand{\intentverifier}{Intent Verifier\xspace}
\newcommand{\llamatouch}{\textsf{LlamaTouch}\xspace}
\newcommand{\challenge}{\textsf{Challenge}\xspace}

\newcommand{\showsolutions}{\long\def\soln##1\solnend{##1}}

\showsolutions

\newcommand{\djl}[1]{{\color{orange}{\soln DJLee: #1\solnend}}}

\newcommand{\kh}[1]{{\color{orange}{\soln KH: #1\solnend}}}

\newcommand{\customunmarkedfootnote}[2]{%
  \begingroup
  \renewcommand\thefootnote{#1}%
  \footnotetext{#2}%
  \addtocounter{footnote}{-1}%
  \endgroup
}

\usepackage{tikz}
\usetikzlibrary{positioning,arrows,calc,automata}

\title[VeriSafe Agent]{VeriSafe Agent: Safeguarding Mobile GUI Agent via Logic-based Action Verification}

\author{Jungjae Lee}
\authornote{Co-first authors : Jungjae Lee, Dongjae Lee}
\affiliation{
  \institution{School of Computing, KAIST}
  \country{Republic of Korea}
}
\email{dlwjdwo00701@kaist.ac.kr}

\author{Dongjae Lee}
\authornotemark[1]
\affiliation{
  \institution{School of Computing, KAIST}
  \country{Republic of Korea}
}
\email{dongjae.lee00@kaist.ac.kr}

\author{Chihun Choi}
\affiliation{
  \institution{Korea University}
  \country{Republic of Korea}
}
\email{clgns0102@korea.ac.kr}

\author{Youngmin Im}
\affiliation{
  \institution{School of Computing, KAIST}
  \country{Republic of Korea}
}
\email{ym.im@kaist.ac.kr}

\author{Jaeyoung Wi}
\affiliation{
  \institution{School of Computing, KAIST}
  \country{Republic of Korea}
}
\email{wijaeyoung@kaist.ac.kr}

\author{Kihong Heo}
\affiliation{
  \institution{School of Computing, KAIST}
  \country{Republic of Korea}
}
\email{kihong.heo@kaist.ac.kr}

\author{Sangeun Oh}
\affiliation{
  \institution{Korea University}
  \country{Republic of Korea}
}
\email{sangeunoh@korea.ac.kr}

\author{Sunjae Lee}
\authornote{Co-corresponding authors : Sunjae Lee, Insik Shin}
\orcid{0000-0002-2755-5511}
\affiliation{
  \institution{Sungkyunkwan University}
  \country{Republic of Korea}
}
\email{sunjae.lee@skku.edu}

\author{Insik Shin}
\authornotemark[2]
\orcid{0000-0002-9128-2415}
\affiliation{%
  \institution{School of Computing, KAIST}
  \country{Republic of Korea}
}
\email{ishin@kaist.ac.kr}

\renewcommand{\shortauthors}{J. Lee, D. Lee, et al.}

\copyrightyear{2025}
\acmYear{2025}
\setcopyright{cc}
\setcctype{by-nc}
\acmConference[ACM MOBICOM '25]{The 31st Annual International Conference on Mobile Computing and Networking}{November 4--8, 2025}{Hong Kong, China}
\acmBooktitle{The 31st Annual International Conference on Mobile Computing and Networking (ACM MOBICOM '25), November 4--8, 2025, Hong Kong, China}\acmDOI{10.1145/3680207.3765248}
\acmISBN{979-8-4007-1129-9/2025/11}

\begin{document}
\begin{abstract}
\customunmarkedfootnote{$*$}{Co-first authors: Junjae Lee, Dongjae Lee.}
\customunmarkedfootnote{$\dagger$}{Co-corresponding authors: Sunjae Lee, Insik Shin.}
	Large Foundation Models (LFMs) have unlocked new possibilities in human-computer interaction, particularly with the rise of mobile Graphical User Interface (GUI) Agents capable of interacting with mobile GUIs. These agents allow users to automate complex mobile tasks through simple natural language instructions. However, the inherent probabilistic nature of LFMs, coupled with the ambiguity and context-dependence of mobile tasks, makes LFM-based automation unreliable and prone to errors.
	To address this critical challenge, we introduce \textsf{VeriSafe Agent (VSA)}\footnote[1]{The system is available at: \nolinkurl{https://github.com/VeriSafeAgent/VeriSafeAgent}}: a formal verification system that serves as a logically grounded safeguard for Mobile GUI Agents. \sys{} deterministically ensures that an agent's actions strictly align with user intent before executing the action. At its core, \sys{} introduces a novel \textit{autoformalization} technique that translates natural language user instructions into a formally verifiable specification.
    This enables runtime, rule-based verification of agent's actions, detecting erroneous actions even before they take effect. To the best of our knowledge, \sys is the first attempt to bring the rigor of formal verification to GUI agents, bridging the gap between LFM-driven actions and formal software verification. 
    We implement \sys using off-the-shelf LFM services (GPT-4o) and evaluate its performance on 300 user instructions across 18 widely used mobile apps. The results demonstrate that VSA achieves 94.33\%--98.33\% accuracy in verifying agent actions, outperforming existing LFM-based verification methods by 30.00\%--16.33\%, and increases the GUI agent's task completion rate by 90\%--130\%.
\end{abstract}

\ccsdesc[500]{Computing methodologies~Mobile agents}
\keywords{GUI Agent, AI Safety; Formal Verification}

\maketitle
\section{Introduction}
The advent of Large Foundation Models (LFMs)~\cite{openai, claude, llama} has revolutionized human-computer interaction, paving the way for a new generation of agents capable of interacting with graphical user interfaces (GUIs)~\cite{autodroid, autodroid2, mobilegpt, appagent, mobileagent, operator, cua, seeact, andworld, cogagent}. Among these, Mobile GUI Agents 
stand out for their ability to automate complex tasks within mobile applications, reducing manual effort and enhancing user convenience.
By leveraging the reasoning and natural language understanding capabilities of LFMs,
these agents can interpret user requests and translate them into sequences of UI interactions~\cite{autodroid, mobilegpt, appagent, mobileagent}.

However, despite significant advancements, existing Mobile GUI Agents still face fundamental limitations that hinder their reliability and safety in real-world applications. 
One major challenge stems from the probabilistic nature of LFMs, which can lead to unpredictable and erroneous actions.
Additionally, mobile app interactions are often ambiguous and context-dependent, making it difficult even for state-of-the-art LFMs to generate consistently accurate actions~\cite{andworld,andarena,llamatouch}. 
These challenges are especially critical for tasks involving sensitive operations, such as financial transactions or private communications, where errors can have serious or irreversible consequences.
Therefore, implementing robust safeguards and verification mechanisms is essential to ensuring the safe and reliable deployment of Mobile GUI Agents.

Recent approaches~\cite{mobileagent,appagent2} have attempted to mitigate these challenges by introducing reflection agents that use LFMs to review the actions of the primary GUI agent and provide feedback.
However, these approaches come with significant drawbacks.
Since reflection agents also rely heavily on LFMs, they remain susceptible to the same errors, leading to a compounding of inaccuracies.
Furthermore, repeated LFM queries for reflection incur substantial computational costs and latency, further limiting their practicality.
These limitations highlight three key challenges inherent in relying solely on LFMs for verification:

\textit{1) Existence of Irreversible and Risky Action:} One way to improve verification accuracy is to analyze both the action and its resulting screen state. While the app screen after the action \textit{(post-action verification)} provides useful context for verification, many critical mobile actions (e.g., making a payment, sending a message) are irreversible. In such cases, verification becomes meaningless, as the agent cannot filter or "undo" the erroneous actions. This fundamentally defeats the core purpose of verification, which prevents critical errors before they occur. To ensure genuine safeguard against critical errors, verification must be performed before executing the action \textit{(pre-action)}.

\textit{2) The Forward Assessment Problem:} To Verify an agent's action before its execution, it requires a forward-looking assessment: predicting the effect of the proposed action before execution. However, this process can be as difficult -- if not more difficult -- than generating the action itself. Furthermore, the predicted outcome aligns with the user's overall goal adds another layer of complexity. Consequently, this approach is akin to solving a simpler problem with a more complex one, potentially turning verification into a performance bottleneck.

\textit{3) Compounding Error Probability:} Mobile tasks typically comprise a sequence of actions, each requiring verification. Continuously relying on LFMs for validation at every step increases the likelihood of accumulated errors, which can ultimately degrade overall system accuracy, despite efforts to improve it. 



To overcome these challenges, we introduce \textsf{VeriSafe Agent (VSA)}, a deterministic, logic-based verification system for Mobile GUI Agents. 
To the best of our knowledge, this is the first attempt to provide a reliable pre-action verification mechanism grounded in logic-based reasoning rather than existing probabilistic methods, bridging the gap between probabilistic actions and deterministic safety.
Specifically, \sys{} leverages the concept of \textit{autoformalization}~\cite{yuhuai2022autoformalization}, which automatically translates natural language user instructions into a formal specification.
Using the translated specification, VSA performs runtime verification to ensure the correctness of the agent's actions in runtime.

However, unlike conventional software verification~\cite{jason2023grounding, ziyi2024plugin, yi2024selp, obi2025safeplan}, which operates on predefined safety requirements, \sys faces the unique challenge of formalizing impromptu, user-defined instructions on-the-fly, across diverse and dynamic landscape of mobile applications.
To accomplish this, \sys{} introduces three key innovations:n 
\begin{enumerate}
    \item \textit{Domain-Specific Language (DSL) and Developer Library\footnote[1]{The library is available at: \nolinkurl{https://github.com/VeriSafeAgent/VeriSafeAgent_Library}}:} A specialized language and accompanying tools tailored to dynamic nature of mobile environments. Collectively, they can encode both the natural-language user instructions and corresponding UI actions as logical formulas.
    \item \textit{Intent Encoder and Verification Engine:} A system that systematically translates user instructions into logical constraints (expressed as our DSL) and performs runtime, deterministic verification against these constraints.
    \item \textit{Structured Feedback Generation:} A proactive mechanism that provides actionable feedback to the GUI agent, \textit{before} executing an action, identifying specific logical violations and unmet conditions to guide the agent toward correct task completion.
\end{enumerate}

We implement a prototype of \sys{} using off-the-shelf online LFM service (GPT-4o) and integrate it with M3A (Multimodal Autonomous Agent for Android) GUI agent~\cite{andworld}. Our evaluation, conducted on 300 user instructions across 18 widely used mobile applications demonstrates that with an estimated $\sim$437 additional lines of code (LoC) per app, \sys{} successfully verifies GUI agent actions with up to 98.33\% accuracy, achieving a false positive rate of 1.3\% and a false negative rate of 0.3\%. This significantly outperforms baseline reflection agents by 23.2\%. Furthermore, \sys{}'s structured feedback, derived from logical verification results, enables the GUI agent to enhance its task completion rate by 130\%, a substantial improvement compared to reflection agents, which failed to correct any tasks.

\section{Background and Motivation}
\label{sec:2}
\begin{figure*}[t]
	\centering
	\includegraphics[width=1.0\textwidth]{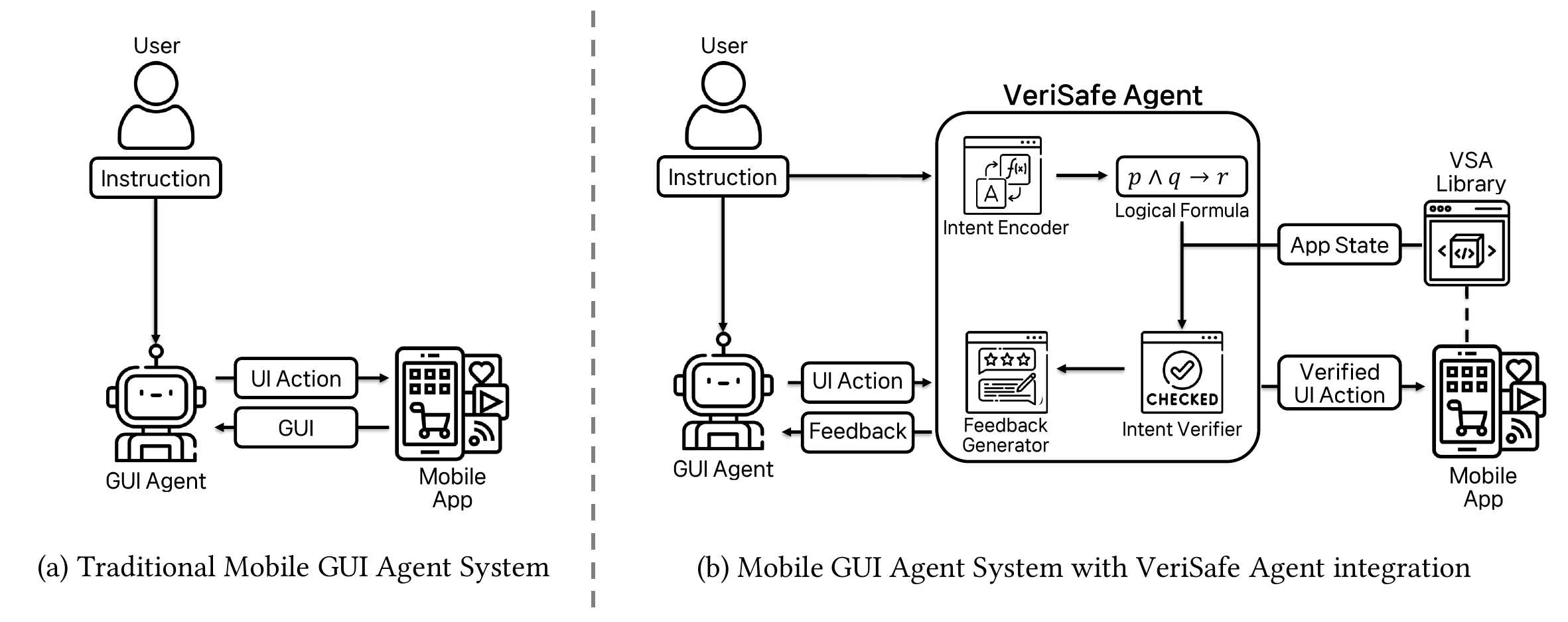}
        \vspace{-0.7cm}
	\caption{Comparison of (a) a traditional GUI agent system and (b) \sys{} integrated system.}
    \vspace{-0.2cm}
	\label{fig:overview}
\end{figure*}

\textbf{Graphical User Interface (GUI) Agents}. GUI Agents~\cite{autodroid, autodroid2, mobilegpt, appagent, mobileagent, operator, cua} are software programs designed to interact autonomously with applications through graphical user interfaces. By interpreting visual and semantic information presented in GUIs and simulating human interactions (e.g., clicks, swipes, and text entry), GUI agents enable automation of digital tasks without requiring internal access to application code or APIs. Recent advancements in artificial intelligence, particularly driven by Large Foundation Models (LFMs), have significantly enhanced the capabilities of GUI agents, enabling them to understand complex user instructions and interpret sophisticated mobile application GUIs. However, due to the inherently probabilistic nature of LFMs and often ambiguous and under-specified nature of GUI interactions, contemporary GUI agents are susceptible to incorrect actions~\cite{andarena,llamatouch,survey1,survey2,survey3,survey4,survey5}, necessitating robust verification mechanisms to ensure reliability and safety in practical scenarios.

\textbf{Safeguarding GUI Agents using Reflection.}
The current practice for safeguarding GUI agents often involves the use of \textit{reflection agents}, which utilize LFMs to review the actions generated by the primary GUI agent. Reflection-based approach can be broadly classified into two categories: pre-action verification and post-action verification.

\textit{Pre-action reflection~\cite{appagent2}} evaluates a proposed action \textit{before} it is executed on the mobile app. This approach serves as an effective guardrail by providing an opportunity to abort or correct erroneous actions before they occur. However, it frequently suffers from low verification accuracy because it requires predicting the outcome of an action. Such inaccuracies lead to high false negatives and positives, paradoxically decreasing the overall system accuracy and undermining the intended safety benefits.

Conversely, \textit{post-action reflection~\cite{mobileagent}} evaluates an action \textit{after} its execution, leveraging the resulting application state to concretely assess the correctness of the action. While this approach significantly improves verification accuracy, it has a critical limitation: it cannot prevent irreversible actions. For irreversible actions such as financial transactions or sending messages, post-action verification becomes ineffective, as such actions cannot be undone once they are performed.

\sys{} is designed to address these limitations, providing a reliable \textit{pre-action} verification that is based on a logic-based deduction instead of probabilistic reasoning.

\textbf{Software Verification for Safety-Critical Systems.} Software verification~\cite{software_verification, model_checking, edmund2018modelchecking, bruno2007staticanalyzer, hardware_verification, autonomous_software_verification, veriphy, modelplex, testing, runtime_verification} encompasses a range of methodologies to ensure that software systems behave correctly according to specification. These formal verification techniques have been widely adopted to provide strong correctness guarantees in safety-critical domains, including hardware design~\cite{hardware_verification}, automotive systems~\cite{autonomous_software_verification}, and cyber-physical systems~\cite{veriphy, modelplex}. Among them, dynamic verification~\cite{testing} aims to check the properties of the program or detect incorrect behaviors when the program is executed. The formal definition of malfunctions is written in mathematical expressions such as first-order logic, linear temporal logic, and finite-state automaton. Once the verification properties are defined, verification can be performed by checking whether an automaton reaches an accept state or by converting the problem into a satisfiability modulo theory (SMT) problem to find satisfying variable assignments using SMT solvers~\cite{z3}. In particular, runtime verification is a type of dynamic verification that monitors the execution of a program in runtime~\cite{runtime_verification}. It detects anomalous behavior through systematic observation of program execution.
\section{\syslong (\textsf{VSA}): Overview}

Inspired by the success of formal verification in other domains, this work adapts techniques from software verification to solve the specific challenges of Mobile GUI Agent safety and reliability.

As illustrated in \autoref{fig:overview}, \sys{} is layered on top of an existing GUI agent, acting as a verification layer \textit{before} the agent's proposed UI actions are injected into a mobile app. Given a user instruction, \sys{} translates it into a logical formula representing the conditions for successful task completion.  Then, when the GUI agent generates a UI action, \sys{} verifies whether this action satisfies the pre-defined logical formula. If the action is verified as correct, it is passed on to the mobile application.  If the action fails verification, \sys{} provides feedback to the GUI agent, explaining the reason for the failure and guiding the agent to generate a corrected action.
\subsection{Challenges}
In doing so, \sys{} addresses three key challenges:
\begin{itemize}
	\vspace{-0.1cm}
	\item [C1.]How to \textit{formally} express the user's intent and the app's execution flow in logically tractable representation?
	\item [C2.] How to \textit{accurately} translate natural language user instructions into a formal representation and perform runtime verification against it?
	\item [C3.] How to \textit{effectively} communicate verification results back to the GUI agent in an actionable, explainable manner?
\end{itemize}
\textbf{C1.}
Conventional software verification techniques typically rely on precise, pre-defined formal specifications of program behaviors. However, mobile GUI agents operate on natural language instructions, which are inherently informal and ambiguous. Therefore, we need a representation system that can clearly express both the user's intent and the corresponding logical flow of the application to fulfill that intent. To address this, \sys{} introduces a domain-specific language (DSL, \autoref{sec:dsl}) and an associated developer library (\autoref{sec:lib}). The DSL provides the syntax and semantics for expressing user intent as logical rules that must be satisfied. The developer library enables app developers to explicitly define the app states and transitions required to construct these rules. Together, they allow \sys{} to represent both the \textit{desired} behavior (user intent) and the \textit{actual} behavior (app execution) in a unified, logically verifiable manner.

\textbf{C2.}
Natural language is often vague, underspecified, and prone to misinterpretation, making it difficult to accurately translate into a formal DSL suitable for logical verification. Even the most capable language models often fail to do this without proper external guidance. To address this challenge, \sys{} employs a \textit{self-corrective encoding} approach (\autoref{sec:intent_encoder}) that enables the language model to iteratively refine its translation based on syntactic and semantic checks. Furthermore, to achieve efficient yet robust verification, \sys{} provides a two-tiered verification strategy (\autoref{sec:intent_verifier}), adjusting verification intensity according to the action's significance.

\textbf{C3.}
Simply flagging an action as incorrect is insufficient for improving the overall practicality of Mobile GUI Agents. To enhance their accuracy and reliability, we must provide actionable feedback that correctly guides agents toward the goal. However, LFM-generated feedback is often vague, underspecified, or even incorrect, potentially leading to repeated errors or suboptimal actions. Therefore, \sys{} incorporates a structured feedback generation mechanism (\autoref{sec:feedback}) that systematically generates precise, rule-based feedback by identifying unmet or violated conditions based on the logical verification results.

\subsection{System Workflow}
\label{sec:workflow}
This section describes the high-level workflow of how \sys{} logically verifies GUI agent actions. 

\textbf{App Development Phase.}
To enable rigorous rule-based verification, \sys{} requires app developers to define the critical states and transitions relevant for verification using \sys{} developer library (\autoref{sec:lib}). This includes declaring application state spaces and their update mechanisms. While this requires additional developer effort, typically involving 5 to 10 lines of code per state, and potentially more depending on the application's complexity, such practices are already common in modern app development workflows~\cite{react,jetpack,swift}.

\textbf{Intent Encoding.} Given a natural language user instruction, the \textit{Intent Encoder} first translates it into structured logical rules capturing essential constraints and logical dependencies within the user instruction.
For example, an instruction \textit{"Book tickets to the movie M at 7 pm"} can be conceptually translated into the following logical rule:
\vspace{-0.1cm}
\begin{equation*}
	\label{eq:example_horn_clauses}
	\textsf{MovieInfo(title=``M'',\,time=7pm)} \rightarrow \textsf{Book}
\end{equation*}
where each constraint (e.g., \textsf{title=``M''}, \textsf{time=7pm}) is a condition must be satisfied before booking the ticket.

\textbf{Verification.}
When the GUI agent generates a UI action, the \sys{} developer library (\autoref{sec:lib}), embedded within the mobile application, intercepts the action and returns the \textit{anticipated} state transition that would result from the execution of this action.
The \textit{Intent Verifier} (\autoref{sec:intent_verifier}) then evaluates whether this state transition is valid according to the logical rules generated earlier. For example, if the state transition indicates that the name of the restaurant is \textsf{`S'} instead of \textsf{`R'}, the \textsf{IsNameR} Boolean variable would be evaluated to \textsf{False}, indicating that the action is invalid and we can't proceed to \textsf{Reserve}.

\textbf{Feedback Generation.} Based on the verification result, \sys{} provides structured feedback to the GUI agent. If verification succeeds, the action is passed on to the mobile app for execution, and \textit{Feedback Generator} (\autoref{sec:feedback}) guides subsequent actions. If verification fails, the action is discarded, and \textit{Feedback Generator} explains precisely which predicates or constraints were unmet, guiding the agent toward generating a correct action.

This workflow continues until either the task reaches a final state or the GUI agent explicitly terminates the task despite feedback, resulting in task failure. Through this systematic approach, \sys{} ensures that Mobile GUI Agents adhere to the logical constraints defined by user instructions while interacting with mobile applications.
\section{Domain-Specific Language}
\label{sec:4}
This section details the domain-specific language (DSL) and the associated developer library. 
DSL and developer library are the basis for formally representing user instructions and GUI agent actions.
Throughout the remainder of this paper, we illustrate \sys{} using the running example: \textit{"Reserve restaurant R before 7 PM. If the restaurant is not available at that time, do nothing."}

\subsection{Design of Domain-Specific Language}\label{sec:dsl}
We first clarify the formal syntax and semantics of our DSL, which enables the formal representation of the user intent and desired app execution flow.
Our DSL is designed to flexibly express a wide range of constraints, execution flows, and conditional branches found in natural language user instructions.

\subsubsection{Syntax and Semantics.}\label{sec:dsl-syntax}
Inspired by logic programming languages (e.g., Datalog~\cite{datalog} and Prolog~\cite{prolog}), our DSL is structured around Horn clauses.
A Horn clause refers to a logical formula with the specific structure $p_1 \land p_2 \land \cdots \land p_n \rightarrow o$, where $p_1, p_2, \ldots, p_n, o$ are predicates evaluating to boolean values.
Horn clauses are optimized for representing objective attainment conditions and execution flow, commonly used in program verification~\cite{jayhorn, seahorn, rusthorn, bjorner2015horn}.
Each Horn clause signifies that the preconditions for $o$ to be satisfied are $p_1, p_2, \ldots, p_n$.
The formal syntax of our DSL is summarized in \autoref{fig:dsl_syntax}.
\begin{figure}[t]
	\begin{bnf}
		\textit{Specification}::= $\textit{Rule} ^ *$;;
		\textit{Rule}::= $\textit{Pred} \land \ldots \land \textit{Pred}\rightarrow$ \textit{Objective}
		| $\textit{Pred} \land \ldots \land \textit{Pred}\rightarrow$ \textsf{Done};;
		\textit{Pred}::= $\textit{Objective} \mid \textit{StatePred}(\textit{Constraint} ^ *)$;;
		\textit{Constraint}::= \textit{Variable}\quad$\textit{Operator}\quad\textit{Constant}$;;
		\textit{Constant}::= $\textit{String} \mid  \textit{Number} \mid  \textit{Boolean}$
		| $\textit{Date} \mid  \textit{Time} \mid  \textit{Enumeration}$;;
		\textit{Operator}::= $= \mid \neq \mid \simeq \mid > \mid \geq \mid < \mid \leq \mid \subseteq \mid \nsubseteq$ ;;
	\end{bnf}
	\caption{Simplified syntax of our DSL}
    \vspace{-0.2cm}
	\label{fig:dsl_syntax}
\end{figure}

\textbf{Specification and Rule.}
In our DSL, a single user instruction is expressed as a set of Horn clauses called \textit{specification}.
Each Horn clause is called a \textit{rule}, which represents a list of predicates (preconditions) and the objective to be achieved (conclusion).
Rules break down a user instruction into multiple smaller steps, making it easier to verify complex instructions incrementally.
Each rule consists of predicates and an objective which represent the precondition and intermediate step, respectively.
Predicates are composed of state predicates and objectives of the other rules.
State predicates specify the conditions that must hold in the application for the intermediate step to be achieved.
When the objective of another rule is used as a predicate, it establishes a precedence relationship between rules.
A rule is satisfied if all predicates hold.
If a rule ends with \textsf{Done} and such a rule is satisfied, the task is considered complete.

\textbf{State Predicate.}
State predicates represent abstract states of applications. They ignore overly specific information (e.g., size of buttons) and represent only the information important for verification.
A state predicate $p$ takes the form $p(c_1, \ldots, c_n)$, where each $c_i$ is a constraint. Each constraint allows us to express a target state precisely. For example, a state predicate of \textsf{RestaurantInfoResult} can have constraints on \textsf{restaurant\_name}, \textsf{location}, and \textsf{cuisine\_type}.

\textbf{Constraint.}
A constraint $c$ has the form $var\; op\; const$ (e.g., $x=10$), where $x$ is a variable representing a specific element within the application state (e.g., \textsf{restaurant\_name}, \textsf{location}).
$op$ is a comparison operator, and $const$ represents a constant value.
The expression $x; op; const$ returns true if the variable $x$ meets the condition defined by the operator $op$ and the constant $const$.
If $x$ has not been observed in the app, it defaults to $\textit{undefined}$, making the constraint always false.
The variable $x$ can be one of the following six types: string, number, Boolean, enumeration, date, and time.
For example, given a constraint $x \geq 10$, if the value of the variable $x$ was observed to be 15 in the application, then the constraint is considered true.

\subsubsection{Running Example.}
\begin{figure}
	\small
	\begin{align*}
		R_1: & \textsf{RestaurantInfo}(\textsf{name}=\textsf{``R''})                                                    \\ &\land  \textsf{ReserveInfo}(\textsf{date}=\textsf{Today}, \textsf{time}<\textsf{19:00}, \textsf{available}=\textsf{True}) \\ &\rightarrow \textsf{Reserve} \\[0.5em]
		R_2: & \textsf{Reserve} \land \textsf{ReserveResult}(\textsf{success}= \textsf{True}) \rightarrow \textsf{Done} \\[0.5em]
		R_3: & \textsf{RestaurantInfo}(\textsf{name}=\textsf{``R''})                                                    \\ & \land\textsf{ReserveInfo}(\textsf{date}= \textsf{Today}, \textsf{time}< \textsf{19:00}, \textsf{available}\neq\textsf{True}) \\& \rightarrow \textsf{Done}
	\end{align*}
    \vspace{-0.7cm}
	\caption{Specification for the instruction \textit{``Reserve restaurant
			R before 7 PM. If the restaurant is not available at that time,
			do nothing.''}}
    \vspace{-0.6cm}
	\label{fig:running-example}
\end{figure}

Consider our running example of reserving a restaurant.
The instruction is translated into the specification shown in \autoref{fig:running-example}.
\textsf{RestaurantInfo} is a predicate that represents conditions on information about the restaurant you want to reserve, with the constraint that the name must be \textsf{``R''}.
\textsf{ReserveInfo} is information about the reservation, with the selected date, time, and availability as constraints.
The first rule, $R_1$, means to make a reservation if the restaurant name is \textsf{``R''} and the reservation is available before \textsf{19:00} today.
The second rule, $R_2$, indicates task completion if the reservation is confirmed to be successful.
The last rule, $R_3$, represents another task completion if the reservation is not available before \textsf{19:00} today.

Notice that the objective \textsf{Reserve} of rule $R_1$ appears as a predicate in rule $R_2$.
This signifies that $R_1$ must be satisfied before proceeding to $R_2$.

Specification is dynamically evaluated during the agent's interaction with the application.
Initially, all variables $x$ are considered \textit{undefined}.
The values of these variables are updated at runtime by the state transitions caused by the GUI agent's actions.
These updates are governed by pre-defined state update methods written in the developer library (see \autoref{sec:lib}).
Whenever a variable's value changes, the associated constraint is re-evaluated.

For example, consider the \textsf{ReserveInfo} predicate. If the GUI agent attempts to reserve restaurant ``\textsf{R}'' at 18:00 today, the update rule will set the values of the \textsf{date} and \textsf{time} variables to \textsf{Today} and \textsf{18:00}, respectively. When we evaluate the constraints against the updated variables, \textsf{date=Today} and \textsf{time<19:00} will both be true. If \textsf{availability} is observed to be true in the app state after selection, all constraints of the \textsf{ReserveInfo} predicate are true, and the predicate itself becomes true. This runtime evaluation of the specification allows \sys{} to continuously monitor the agent's progress and verify whether its actions are leading the application toward a state that satisfies the user's intent.

\subsection{Developer Library}
\label{sec:lib}
To effectively generate \textit{specification} from a user instruction and evaluate it against agent's actions, \sys{} requires a definition of two key components: \textit{(i)} the set of candidate state predicates (i.e., the application’s state space) that serve as building blocks for the specification, and \textit{(ii)} the mechanisms through which each predicate’s constraint variables $x$ are updated during app execution (i.e., state transitions).

To this end, \sys{} provides a \textit{developer library} that empowers app developers to explicitly define application states (as state predicates) and their corresponding update mechanisms (as state transitions). This approach capitalizes on developers' deep understanding of their application's internal logic and intended behavior. Although it requires additional developer effort, it enables highly accurate state definitions suitable for rule-based logical verification---a level of precision that is unattainable with automated approaches like static code analysis or LFM-driven dynamic analysis.

Furthermore, this approach aligns with established practices in modern mobile app development.
Frameworks like React Native (state management~\cite{react}), Android's Jetpack Compose (state hoisting~\cite{jetpack}), and iOS's SwiftUI (\texttt{@State}~\cite{swift}) all encourage developers to manage app state and transitions.
By closely mirroring these familiar paradigms, \sys{} ensures seamless, intuitive, and minimally invasive integration into standard app development workflows.

\subsubsection{States Definition.}
A \textit{state}, in the context of the \sys{} Library, is characterized by a set of typed variables (as defined in \autoref{sec:dsl-syntax}) that capture essential details relevant to the application's current context---each variable corresponds to the constraint of the state predicate. Each state definition includes \textit{i) name, ii) description, and iii) variables}:

Developers can define these states using an external JSON file or provided APIs (i.e., \textsf{defineState()}):

\begin{lstlisting}[language=json, basicstyle=\footnotesize \ttfamily, breaklines=true, showstringspaces=false, frame=none]
  {
    "name": "RestaurantInfo",
    "description": "Information about the restaurant you want to reserve.",
    "variables": [{"name": "String"}]
  }
\end{lstlisting}
\noindent
These defined states are then used as candidate state predicates when \sys{}'s Intent Encoder (\autoref{sec:intent_encoder}) translates user instruction into a specification.

\subsubsection{Pre-action State Update Triggers.}
\label{sec:trigger}
\sys{} enables developers to define \textit{triggers} specifying when and how state variables are updated.
Triggers correspond directly to meaningful state transitions within the mobile application.
Most triggers are associated with UI interaction handlers (e.g., \texttt{onClickListener}) or critical points in the application logic.

Triggers return the \textit{expected} state update before executing the action.
The returned update is used to verify the action.
This ahead-of-time update provides two key benefits: \textit{i)} It prevents irreversible actions (e.g., financial transaction, sending a message) from being executed if they violate the user's intent; \textit{ii)} It allows the GUI agent to easily retry with a different action without having to revert the effects of a previous, incorrect action (e.g., deleting incorrect text input).

To facilitate pre-action updates, the \sys{} developer library provides wrapper listeners for common UI input handlers.
These wrappers intercept user interactions and update the relevant state variables according to the developer-defined logic \textit{before} executing the original event handler's code.
The original \texttt{`onClick'} code which performs the actual operation is executed only if the verification succeeds.
For example, consider the \textsf{`RestaurantInfo'} state predicate defined earlier. A developer could associate a trigger with the search button's click listener as follows:
\lstset{
	tabsize = 2, 
	showstringspaces = false, 
	commentstyle = \color{gray}, 
	keywordstyle = \color{blue}, 
	stringstyle = \color{red}, 
	rulecolor = \color{black}, 
	basicstyle = \footnotesize \ttfamily , 
	breaklines = true, 
	numberstyle = \tiny,
}
\vspace{-0.15cm}
\begin{lstlisting}[language = Java , firstnumber = last , escapeinside={(*@}{@*)}]
searchButton.VSAOnClickListener(() -> {
  vsaStateManager.updateState("RestaurantInfo", {
      "name": searchTextField.getText(),
    });
    /* existing code for onClickListener */
  });
\end{lstlisting}

Another key advantage of developer-defined state predicates is the control over \textit{granularity}.
In many mobile tasks, individual atomic actions are not inherently meaningful.
For example, in an e-commerce application, entering individual fields of a shipping address (street, city, zip code) is a set of preparatory steps.
The critical state change is the \textit{submission} of the complete address.
Verifying every single atomic action would be inefficient and could mislead LFMs to think that some steps are incomplete.
By allowing developers to strategically define verification checkpoints that have a meaningful impact on the task's execution, \sys{} can significantly improve the efficiency and accuracy of task verification.

\subsubsection{Post-action State Update Triggers.}
While pre-action verification is ideal for preventing errors, some action outcomes are inherently unpredictable (e.g., the result of a network request). For these scenarios, \sys{} supports \textit{post-action} state updates. Developers can call the \textsf{`updateState()'} API directly, often within asynchronous callbacks, to update the state after the outcome of an action is known. This allows developers to handle dynamic application behaviors and real-world uncertainties such as network delays, asynchronous operations, error conditions, and dynamically determined content.

For instance, consider a flight booking app where the price and availability of flights change dynamically based on real-time data. A developer can define a \textsf{FlightSearchResult} state with variables for price, availability, and departure time. They can then associate a state update trigger with the asynchronous network callback that gets executed when the user clicks the “Search” button. Upon receiving the network response, the developer-defined trigger logic updates the \textsf{FlightSearchResult} based on the actual results. If the network fails or no seats are available, the trigger can handle these conditions appropriately, updating the state to reflect the error or unavailability. This ability to embed complex, conditional logic directly into the verification framework enables \sys{} to handle a wide range of user requirements and edge cases.

\section{Verification Engine}
\label{sec:verification_engine}

\subsection{Intent Encoder}
\label{sec:intent_encoder}
Natural language instructions cannot be directly used for formal verification; a formalization process is required.
We optimize the LFM-based autoformalization method in the context of mobile agents to automatically convert instructions into specifications written in our DSL.
Relying solely on LFM's output poses risks due to their probabilistic nature and potential for hallucinations.
To address this limitation, we applied \textit{self-corrective encoding}, which corrects errors automatically, and \textit{experience-driven encoding} techniques that gradually adapt to each app.

\subsubsection{Self-Corrective Encoding.}
\label{sec:self-corrective_encoding}
To ensure the encoded specification is correct and includes all relevant constraints, \sys{} employs two complementary self-corrective techniques: (i) rule-based syntax checking, and (ii) decoding-based semantics checking. These methods allow the LFM to incrementally identify, correct, and refine its translations, improving accuracy and reliability.

As an initial step, the \textit{Intent Encoder} prompts the LFM with: i) the user's natural language instruction and ii) the set of developer-defined states (from the developer library). From this information and user instruction, the Intent Encoder generates a draft specification representing the user's intent using available states. The draft specification then undergoes two checking stages:

\textbf{Rule-based Syntax Checking.}
Inspired by conventional programming languages,~\sys{} checks that the generated Specification strictly adheres to the DSL's syntax (\autoref{sec:dsl-syntax}) and that the types of constants $c$ used within the Specification are consistent with the types of variables declared in the developer-defined states.
For example, if the LFM generates a state predicate like
$\textsf{RestaurantInfo}(\textsf{name}\geq\textsf{100})$ while the type of the name variable is defined as `String' in the developer library, this inconsistency is immediately flagged as a syntax error. If an error is detected, a structured error message is returned to the encoding LFM as feedback, guiding it toward a corrected translation.

\textbf{Decoding-Based Semantics Checking.}
To further validate the semantic correctness of the translation, the generated Specification is \textit{decoded} back into a natural language description using a separate \textit{decoder LFM}. A \textit{checker LFM} then compares this decoded description to the original user instruction, confirming that the encoding genuinely captures the user's intent rather than merely manipulating symbols. If discrepancies are found, the checker LFM identifies the root cause and provides feedback to the \textit{encoding LFM}.

These complementary checking mechanisms significantly improve the robustness and accuracy of the intent encoding process, effectively mitigating risks associated with solely relying on probabilistic LFM outputs.

\subsubsection{Experience-driven Encoding.}
\label{sec:experience-driven_encoding}
Although \textit{Self-corrective Encoding} significantly enhances translation accuracy, it occasionally produces hallucinations when presented with an enormous number of candidate state predicates. 

To address this challenge, \sys{} introduces an \textit{Experience-driven Encoding}, which leverages previously successful encodings to guide subsequent translations.
When an instruction is successfully translated and verified, its result specification is cached.
Since a single specification consists of multiple rules representing an instruction’s intermediate objectives, numerous instructions for the same application may share similar rules.
By retrieving and prioritizing predicates based on past encoding experience, we can significantly reduce the search space for the LFM encoder, improving both efficiency and consistency.

\subsection{Intent Verifier}\label{sec:intent_verifier}
The verification process is straightforward. When the GUI agent generates an action, the associated state updates (defined via the developer library, \autoref{sec:lib}) are triggered. They update the values of the variables $x$ of the corresponding constraints and state predicates in the Specification. After each update, the affected predicates are re-evaluated to a new truth value.

Based on this basic mechanism, \sys{} performs two levels of verification: \textit{Predicate-level Verification} at the predicate level and \textit{Rule-level Verification} at the rule level.

\textbf{Predicate-level Verification} is performed at \textit{every} state update. Whenever a state predicate within the Specification is re-evaluated due to a variable update, its truth value is checked. If a predicate evaluates to false, this indicates a \textit{potential} violation of the user's intent. For example, if the Specification includes the predicate $\textsf{RestaurantInfo}(\textsf{name}=\textsf{``R''})$, and the agent's action causes the \textsf{name} variable to be updated to \textsf{``S''}, the predicate would evaluate to false, indicating a \textit{potential} violation of user intent.

It is important to note that an incorrect predicate at this stage does not necessarily mean a definitive error because the updated state may represent an intermediate step necessary to eventually reach the final, correct state.
For example, when buying three apples, the number of apples might be incremented one at a time. Each intermediate update (1 apple, then 2 apples) would result in a false value for a predicate requiring three apples, until the final update.

Therefore, upon detecting a failed verification, \sys{} reverts the predicate update and provides the GUI agent with \textit{soft feedback} (\autoref{sec:feedback}).
This feedback serves as a warning, indicating a potential deviation from the intended path, and encourages the agent to double-check its subsequent actions.

\textbf{Rule-level Verification} is performed at irreversible or critical checkpoints within the task execution. To identify these checkpoints, we modified the GUI agent's prompt so that it generates a \textit{critical} flag along with the action if that action is intended to directly achieve one of the objectives $o$ within the Specification (e.g., \textsf{Reserve}).

When the GUI agent generates a \textit{critical} action, \sys{} checks for the complete satisfaction of the conditions required to achieve a specific Objective $o$. If any predicate in the corresponding rule is unsatisfied, the action is considered invalid and blocked. Hard feedback is provided to the agent, detailing the specific unmet conditions. This prevents the execution of potentially irreversible or incorrect actions that would definitively violate the user's intent.

This two-tiered verification approach guides the agent toward its objectives while ensuring the safe execution of high-risk actions.

\section{Structured Feedback Generator}
\begin{figure}
    
    \begin{flushleft}
        {\small        
            \noindent
            \textbf{$\mathbf{F_1}$}: \textit{"To perform \textsf{Reserve}, 1. RestaurantInfo that represents Information about the restaurant you want to reserve should have `name' equal to ``R''; 2. ReserveInfo that represents reservation details should have `date' equal to ``Today'' and `time' less than 19:00, and `available' equal to True. So far, you have achieved step 1."}
            \vspace{0.8em}
            
            \noindent
            \textbf{$\mathbf{F_2}$}: \textit{"To complete the task, 1. perform \textsf{Reserve} 2. ReserveResult that represents reservation result should have `success' equal to "True"."}
            \vspace{0.8em}

            \noindent
            \textbf{$\mathbf{F_3}$}: \textit{"To complete the task, 1. RestaurantInfo that represents Information about the restaurant you want to reserve should have `name' equal to ``R''; 2. ReserveInfo that represents reservation details should have `date' equal to ``Today'' and `time' less than 19:00, but `available' not equal to True. So far, you have achieved step 1."}
        }
    \vspace{-0.2cm}
    \caption{An example of roadmap feedback. Each $F_i$ corresponds to Rule $R_i$ in the Specification}
    \label{fig:roadmap}
    \vspace{-0.5cm}
    \end{flushleft}
\end{figure}

\label{sec:feedback}
The effectiveness of a verification system hinges not only on error detection but also on its ability to guide the agent towards correct behavior. This section explains how \sys{} generates structured feedback based on verification results. \sys{} provides three distinct types of feedback: i) Roadmap Feedback, ii) Predicate-level Soft Feedback, and iii) Rule-level Hard Feedback. 

\subparagraph{\textbf{Roadmap Feedback.}}
\sys{} provides roadmap feedback that outlines the overall path toward task completion regardless of verification outcome.
Because the logical Specification of \sys{} explicitly encodes the preconditions required for achieving the user's objective, it naturally serves as a comprehensive guideline for the GUI agent. 

Specifically, the Feedback Generator converts each rule in the Specification into a natural language explanation, detailing which predicates must be satisfied for each objective. These explanations incorporate developer-defined descriptions for each state predicate, constraints, and currently satisfied predicates. For example, \autoref{fig:roadmap} demonstrates the roadmap feedback for the restaurant reservation scenario after successfully searching for restaurant ``R''.

\noindent
Unlike approaches relying on ambiguous action histories or abstract planning, this structured guidance clearly communicates both current progress and next objectives.

\textbf{Predicate-level Soft Feedback.}
When predicate-level verification fails (i.e., a state predicate evaluates to false), \sys{} provides feedback indicating that the agent's action \textit{may} be incorrect. This feedback includes a description of the desired state for the violated predicate, allowing the agent to reconsider its action. Because predicate-level errors could represent intermediate steps rather than genuine errors (as described in \autoref{sec:intent_verifier}), soft feedback is advisory rather than strictly prohibitive. If the GUI agent generates the same action even after receiving the feedback, \sys{} permits the action, acknowledging its non-critical nature.

\textbf{Rule-level Hard Feedback.}
For critical actions subject to rule-level verification, \sys{} provides comprehensive feedback at the rule level, detailing all state predicates that remain unfulfilled. Unlike soft feedback, if the GUI agent generates the same action again after receiving hard feedback, \sys{} will not allow the action to proceed. Instead, it enforces that all required state predicates must be satisfied before permitting the critical action to be executed.

Through these structured, deterministic feedback mechanism, \sys{} effectively guide GUI agents toward correct task completion, significantly outperforming existing reflection-based methods. By eliminating ambiguity and providing clear, actionable guidance grounded in formal logic, \sys{} enhances the reliability, safety, and practicality of automated mobile task execution.

\section{Implementation}
\begin{figure*}[t]
	\centering
	\includegraphics[width=1\textwidth]{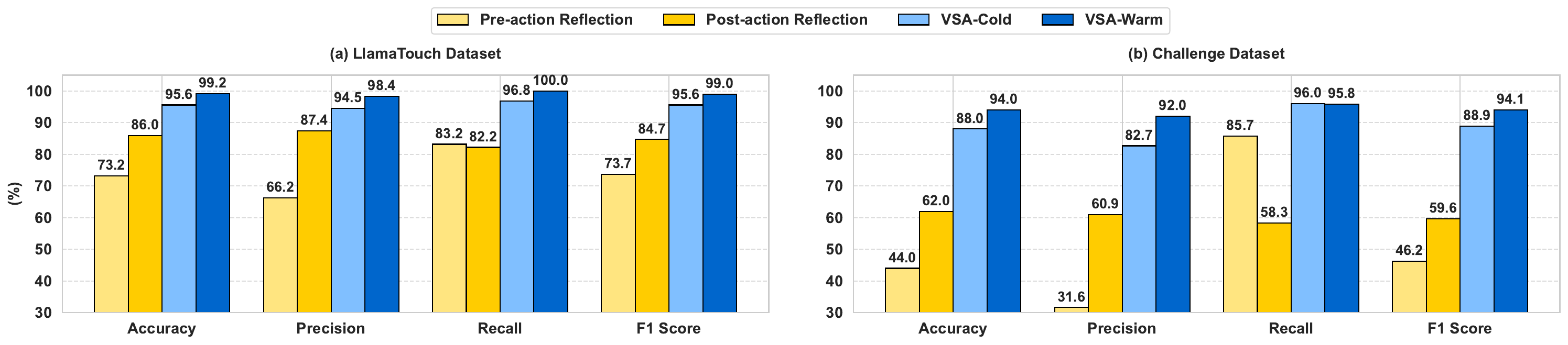}
        \vspace{-0.7cm}
	\caption{Verification accuracy}
        \vspace{-0.2cm}
	\label{fig:accuracy}
\end{figure*}

Our \sys implementation is designed as a modular component that can be seamlessly integrated with existing GUI agents without requiring modifications to their underlying architecture. \sys is implemented using Python, and its Intent Encoder leverages off-the-shelf LFM GPT-4o.

\textbf{Developer Library Guidelines.}
Developers should carefully determine the granularity of State Predicates to suit their application context.
Overly fine-grained State Predicates can make the encoded specification too detailed, increasing the false positive rate.
Conversely, coarse-grained State Predicates may introduce loopholes in the encoded specification, leading to a higher false negative rate.
To assist with finding the sweet spot, the \sys Library provides detailed documentation and guidelines on how to effectively define State Predicates. Given the potential severity of false negatives, it's recommended to begin with a fine-grained verification first and then iteratively refine the predicates to minimize false positives.

\textbf{String Similarity.}
To compare strings semantically rather than structurally, we use OpenAI’s text-embedding-3-small model~\cite{openai} and compute cosine similarity. Based on empirical studies, we set a similarity threshold of 0.7 to determine equivalence.



\section{Evaluation}

In this section, we evaluate the performance of \sys{} against baseline guardrail frameworks in terms of \textit{i) verification accuracy}, \textit{ii) feedback effect}, and \textit{iii) latency and cost}. Throughout the experiments, we integrated each framework, including \sys{}with the \textit{M3A} mobile GUI agent~\cite{andworld}---a simple yet powerful mobile GUI agent powered by GPT-4o---and conducted evaluations using a Google Pixel 8 smartphone.

\textbf{Baselines \& \sys Variants.} To evaluate \sys{}, we compare it with two LFM-based reflection schemes: Pre-action reflection~\cite{appagent2} and Post-action reflection~\cite{mobileagent}, both of which utilize GPT-4o as their backbone LFM. For \sys{}, we evaluate two variants: \sys{}-cold and \sys{}-warm. \sys{}-cold encodes user instructions solely using self-corrective encoding, without cached memory. On the other hand, \sys{}-warm leverages both self-corrective encoding and experience-driven encoding, assuming that a cached list of candidate predicates is available for the given instruction (see \autoref{sec:intent_encoder} for details).

\subsection{Dataset}
To comprehensively evaluate \sys{}, we constructed a dataset\footnote[1]{The dataset is available at: \nolinkurl{https://github.com/VeriSafeAgent/VeriSafeAgent}} of 300 user instructions, divided into two complementary subsets of 150 instructions each: the \textit{Correct} dataset and the \textit{Wrong} dataset.

Each instruction is paired with an execution path---a sequence of UI actions performed by the GUI agent. In the \textit{Correct} dataset, instructions are matched with execution paths that accurately fulfill the user's intent. In contrast, the \textit{Wrong} dataset deliberately introduces mismatches between instructions and their corresponding execution paths. Specifically, we altered key details in the original \textit{Correct} instructions to create discrepancies. For example, \textit{"On Play Books, turn to the search page and check for top-selling books"} was modified to \textit{"On Play Books, turn to the search page and check for newly released books"}, making the original execution path no longer satisfy the updated instruction. Such cases test the verification framework’s ability to detect errors, enabling evaluation of both True Positives (correctly rejecting invalid executions) and False Negatives (incorrectly accepting invalid executions).

Of the 150 unique instructions, 125 were sampled from the \llamatouch dataset~\cite{llamatouch}, selected based on app popularity (global usage), instruction realism (likelihood of real-world occurrence), and current app availability. We excluded tasks that were overly localized, deprecated, or overly specific and prone to overfitting. The remaining 25 instructions were newly created to form the \textit{Challenge} dataset, designed to push the system to its limits with longer action sequences, multiple constraints, and conditional branches. Spanning 9 different applications, these instructions average 19.16 steps (maximum 40), far exceeding \llamatouch's average of 5.67 steps.

Finally, as described in \autoref{sec:lib}, \sys{} relies on developer-defined state specifications and state-update triggers. To emulate this in closed-source mobile apps, we manually annotated state predicates and state transitions for each GUI screen and action in the dataset. This manual annotation serves as a surrogate for direct integration with the developer library, effectively replicating the library's intended functionality.

\subsection{Verification Accuracy} \label{sec:verification}
\autoref{fig:accuracy} presents the verification accuracy of different verification methods, evaluated using standard metrics: Accuracy, Precision, Recall, and F1 Score. Notably, \sys{} consistently outperforms reflection-based methods across all metrics and datasets, achieving near-perfect accuracy on the \llamatouch dataset. 

For simple tasks in \llamatouch dataset, post-action reflection achieves significantly higher accuracy compared to pre-action reflection. This is because it leverages the action's resulting screen as additional context for verification. However, it has critical limitations: it cannot correct irreversible actions. Moreover, our results indicate that this advantage diminishes significantly as instructions become more complex and involve longer action sequences (as seen in the Challenge dataset). This trend highlights the compounding effect of errors stemming from over-reliance on LLMs for verification.

\begin{table}[t]
\centering
\resizebox{0.47\textwidth}{!}{%
  \begin{tabular}{c|c|c|c}
    \toprule
    \textbf{Error Type} & \textbf{Variant} & \textbf{LlamaTouch} & \textbf{Challenge} \\
    \midrule
    \multirow{2}{*}{Overall Missing Predicate} 
      & \sys-Cold & 76/250 & 24/50 \\
      & \sys-Warm & 24/250 & 14/50 \\ 
    \midrule
    \multirow{2}{*}{Critical Missing Predicate}
      & \sys-Cold & 10/250 & 3/50 \\
      & \sys-Warm & 3/250 & 2/50 \\
    \midrule
    \multirow{2}{*}{Superfluous Predicate} 
      & \sys-Cold & 7/250 & 4/50 \\
      & \sys-Warm & 0/250 & 0/50 \\ 
    \midrule
    \multirow{2}{*}{Constraint Mismatch} 
      & \sys-Cold & 8/250 & 11/50 \\
      & \sys-Warm & 4/250 & 6/50 \\
    \bottomrule
  \end{tabular}%
}
\caption{\sys predicate translation accuracy}
\vspace{-1.0cm}
\label{tab:predicate-accuracy}
\end{table}

In contrast, \sys{} performs \textit{pre-action} verification while maintaining high performance across both datasets, thanks to its design that encodes the specification once and performs subsequent verification in a rule-based manner, regardless of task complexity or length. Notably, \sys{}-Warm achieves near-perfect results on \llamatouch and significantly outperforms all baselines on \challenge, highlighting the benefit of a well-constructed predicate memory. While building such a memory (e.g., via offline exploration, user demonstrations, runtime analysis) is beyond this paper’s scope (\autoref{sec:discussion}), these results underscore its promise.

An interesting observation is the high recall of Pre-action Reflection on the \challenge Dataset. However, this comes at the cost of a high false positive rate (72\%), leading to excessive error flagging when instructions become complex. While this increases the chance of detecting \textit{any} error, the likelihood that a flagged action is truly erroneous remains low (low precision).


\textbf{Predicate Translation Accuracy.}
To gain deeper insight into the verification results, we further evaluated \sys{} at the state predicate level. We manually constructed a ground-truth set of predicates $P_{GT}$ for each instruction and compared it against $P_{Spec}$ generated by \sys{}.

\sys's Verification errors fall into three categories:
\textit{1)} \textbf{Missing Predicate} — necessary predicates absent from the specification, potentially allowing incorrect behaviors (False Negatives);
\textit{2)} \textbf{Superfluous Predicate} — unnecessary predicates included, potentially rejecting correct behaviors (False Positives);
\textit{3)} \textbf{Constraint Mismatch} — correct predicate but with inaccurate constraints, which can cause both False Negatives and False Positives.

\autoref{tab:predicate-accuracy} shows the counts of such errors. Note that  \textit{Overall Missing Predicate} counts all predicates absent relative to $P_{GT}$, which defines the maximal set of possible predicates (including semantically overlapping ones). In practice, \sys{} tends to select a minimal set of core constraints, which can lead to an overestimation of “missing” predicates. We therefore also report \textit{Critical Missing Predicates}, which are those whose absence directly affects verification results.

A key observation is that \sys-Warm significantly reduces both missing and superfluous predicates. This demonstrates that predicate memory effectively improves translation accuracy by narrowing the search space for necessary predicates. Furthermore, the distribution of errors across all three types indicates that, despite \sys's strong overall verification performance, residual risks from predicate translation remain. In the Discussion section, we propose complementary policies to mitigate these risks and further enhance \sys's safety.

\textbf{Randomness Mitigation.} 
Although \sys{} issues only a single LFM query per task, repeated verification of the same instruction may yield inconsistent results due to the stochasticity of LFMs. The same inconsistency applies to all methods that invoke an LFM, including reflection-based approaches. To mitigate this randomness, we employ a majority-voting scheme that repeats verification $N$ times and returns the majority decision (\textit{Majority-at-$N$}).

\begin{figure}[t]
	\centering
	\includegraphics[width=0.5\textwidth]{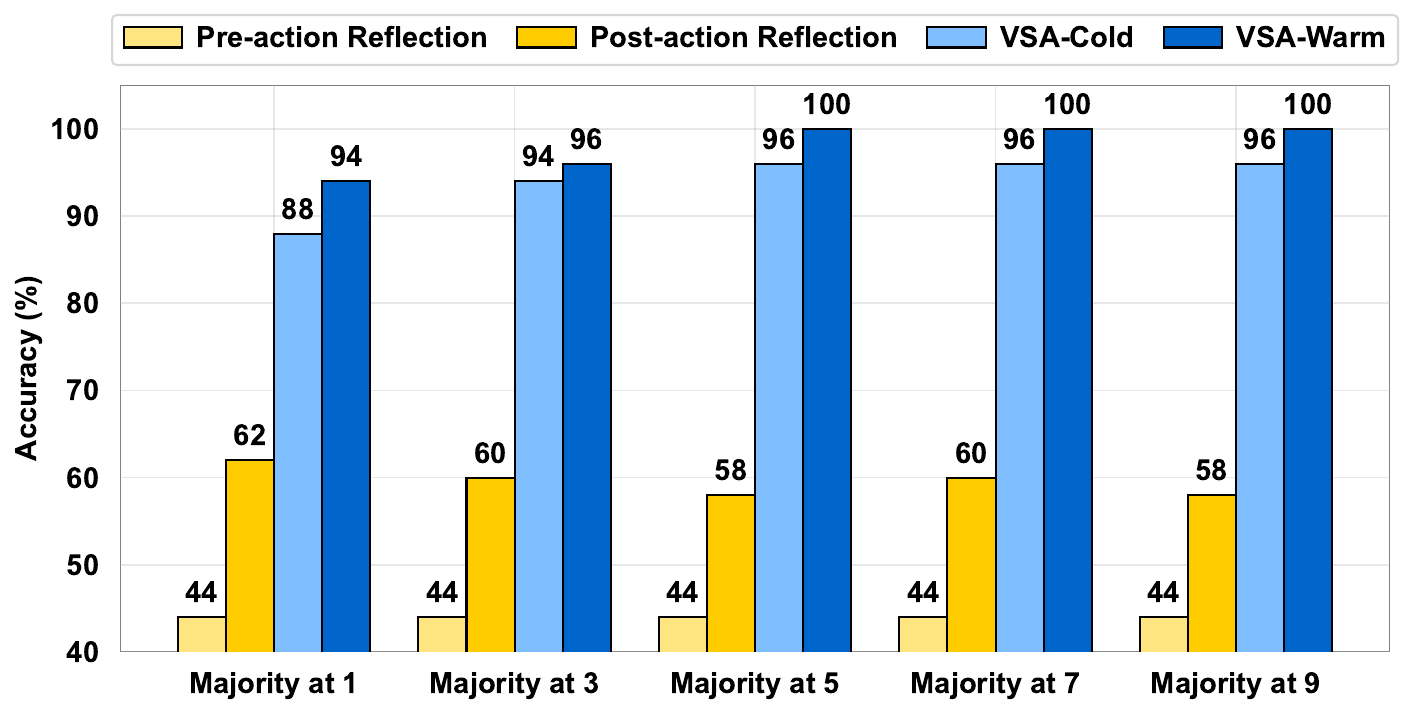}
    \vspace{-0.7cm}
	\caption{Verification accuracy with \textit{Majority-at-N}}
    \vspace{-0.5cm}
        \label{fig:majority}
\end{figure}

\autoref{fig:majority} reports accuracy on 50 instructions from the \challenge dataset as $N$ increases. Reflection-based schemes show limited gains and remain less accurate overall, whereas both \syscold and \syswarm improve substantially, reaching 96\% and 100\%, respectively.

This performance disparity stems from differences in verification mechanisms: reflection schemes produce direct true/false judgments with relatively low variance, whereas \sys{} encodes a logical specification with multiple predicates---small stochastic differences can flip individual predicate decisions---making majority voting especially effective at averaging out noise.

These findings indicate that \sys can achieve substantial performance gains through majority voting. However, increasing verification iterations introduces considerable overhead in terms of latency and cost. We analyze this trade-off in detail in \autoref{sec:overhead}.

\subsection{Effectiveness of Feedback Generation }
The usefulness of a verification system lies not just in identifying errors, but in improving the GUI agent's performance through feedback generation. This section evaluates \sys's logic-based feedback generation against the LLM-based feedback used in reflection methods.

We instructed the M3A GUI agent to execute 25 challenging instructions from our \challenge dataset. Initially, without external feedback, M3A successfully completed only 10 of 25 tasks (40\% success rate). We then integrated each verification system with M3A and re-executed the failed instructions to observe whether the feedback can correct the previously failed tasks.

\begin{figure}[t]
	\centering
	\includegraphics[width=0.5\textwidth]{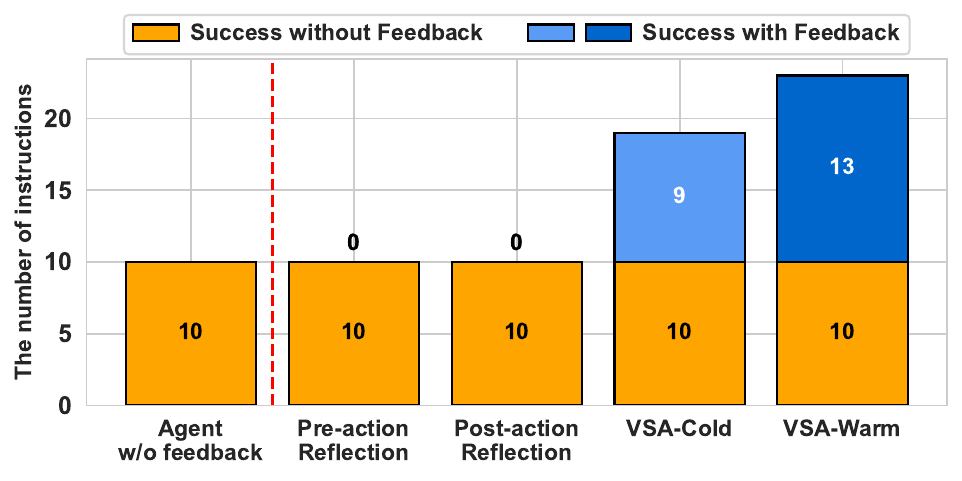}
    \vspace{-0.7cm}
	\caption{Effect of feedback on task completion}
    \vspace{-0.5cm}
        \label{fig:feedback}
\end{figure}

\autoref{fig:feedback} summarizes the results. Notably, Neither pre-action nor post-action reflection corrected any previously failed tasks. In contrast, \syscold{} corrected 9 of 15 failures (60\%), and \syswarm{} corrected 13 of 15 (86\%). The two remaining failures under \syswarm{} were due to unconventional app designs rather than misunderstanding of context. For instance, one clock application required time input as a single numerical sequence (e.g., "143000" for 14:30:00) rather than separate hour, minute, and second components. Such unconventional interfaces could limit \sys's ability to provide meaningful feedback. Nevertheless, given that M3A employs GPT-4o---one of the most capable LLMs---achieving this performance improvement through systematic harness demonstrates \sys's effectiveness.

\sys's superior performance stems from its ability to generate precise, deterministic feedback that clearly identifies what needs correction (i.e., unmet predicates). Moreover, \sys's roadmap feedback proactively guides agents toward correct execution paths, preventing errors from happening in the first place. For example, when asked to upvote the most controversial comment in a subreddit, M3A without feedback repeatedly toggled the upvote button because it could not identify whether the upvote has been properly set. However, when \sys provided clear guidance that the precondition to set 'upvote' to 'true' had been satisfied, M3A completed the task without unnecessary repetition.

\begin{figure*}[t]
	\centering
	\includegraphics[width=1\textwidth]{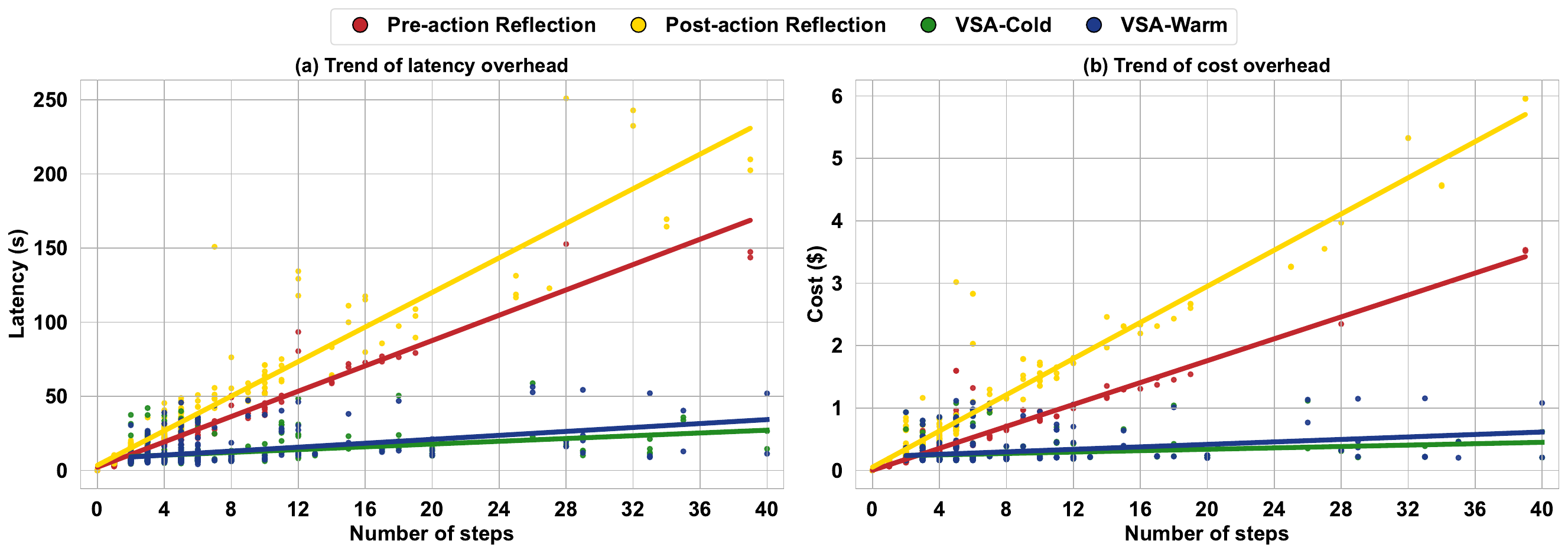}
    \vspace{-0.7cm}
	\caption{Latency and cost overhead of verification methods with respect to the number of action steps.}
    \vspace{-0.2cm}
        \label{fig:latency_cost on task completion}
\end{figure*}

By contrast, reflection methods rely on LFM-generated narrative feedback that is often misleading or inconsistent; we observed cases where a reflection method initially flagged an action as incorrect but reversed its judgment when the agent repeated the same action, indicating indecisive and unreliable guidance.

A potential limitation of \sys{}, though not observed in our evaluation, is that false positives could prematurely terminate tasks that the agent might otherwise complete successfully. Unlike reflection-based methods, where agents typically recover within 1--3 actions after misleading feedback, a single incorrect predicate in \sys's specification could halt task execution entirely. However, our evaluation demonstrates that \sys's decisive feedback provides practical benefits that significantly outweigh the risks of rare false positives. In safety-critical applications such as automated mobile tasks, preventing erroneous actions—even at the cost of occasional false positives—is preferable to allowing potentially harmful actions to proceed.

\subsection{Latency and Cost} \label{sec:overhead}
\sys not only enhances verification reliability but also significantly reduces latency and cost overhead. \autoref{fig:latency_cost on task completion} plots the additional latency and cost incurred by verification as a function of task length (number of steps). As shown by the fitted lines, \sys consistently maintains lower overhead than reflection-based methods across all task lengths. Notably, because reflection-based approach requires an additional LFM query for each action, its overhead scales proportionally with task length. While not depicted in the graph for clarity, its overhead is comparable to that of the GUI agent itself, effectively doubling the total expense.

In contrast, \sys shows near constant overhead, independent of task length. This is because \sys{} uses LFM only once -- when encoding user instructions into formal specifications. All subsequent action verifications are performed using rule-based logic. Although \autoref{fig:latency_cost on task completion} shows a slight upward trend in \sys's overhead for longer tasks, this increase is attributed to more complex instructions  naturally producing longer specifications with additional constraints. However, this increase remains trivial compared to the $O(N)$ complexity of reflection-based methods.

GUI Agents inherently incur substantial latency and cost due to their reliance on complex cognitive reasoning. As shown in \autoref{fig:latency_cost on task completion}, tasks exceeding 10 steps can cost over \$1 and take nearly a minute to complete. Therefore, minimizing this overhead is as critical for practical deployment as achieving high accuracy. \sys thus proves to be more suitable and practical for real-world deployment across all dimensions compared to existing approaches.

\subsection{Developer Burden}
To estimate the developer effort required for VSA integration, we analyzed 18 commercial mobile applications from our dataset. Since these applications are closed-source, we simulated the integration process by manually enumerating the verification-relevant state predicates for each of the 18 target apps. We then estimated the lines of code (LoC) by multiplying predicate counts by the average LoC per predicate measured from our reference implementation.

On average, each application required 58.77 predicates, translating to approximately $\sim$437 LoC. Across all 18 apps, we generated 1,058 predicates in total, of which 540 (51\%) were activated during our experiments. The number of predicates varied with application complexity. Google Tasks, a single-function app, required only 22 predicates, while feature-rich applications like Instagram, X (Twitter), and Facebook required 81, 91, and 72 predicates, respectively. This correlation between application complexity and predicate count is expected, as more complex applications naturally have more states to verify.

While these estimates suggest a non-negligible implementation effort, several factors mitigate the burden. First, many predicates follow standard patterns that could be templated or automated. Second, the predicate definitions align naturally with existing application state logic that developers already maintain. Third, predicate granularity provides a practical lever for balancing coverage and effort, allowing developers to selectively implement only safety-critical predicates.

Overall, the effort required for predicate definition represents a one-time investment that yields ongoing benefits in error prevention and agent reliability. For applications where automated agents perform high-stakes tasks, the safety improvements demonstrated in our experiments justify this initial development effort.

\section{Related Work}

\textbf{Validating LFM responses.}
The probabilistic nature of LFMs often leads to hallucinated or inaccurate content. Numerous methods have been proposed to mitigate these issues, including Chain-of-Verification (CoVe)~\cite{dhuliawala2023chain}, concept-level validation and rectification (EVER)~\cite{kang2023ever}, self-refinement ~\cite{madaan2023selfrefine}, and retrieval-augmented knowledge grounding (Re-KGR) ~\cite{niu2024mitigating}. These approaches typically leverage additional LFM invocations for self-reflections~\cite{appagent2,mobileagent} or rely on external validation against ground-truth data or knowledge bases ~\cite{min2023factscore, chern2023factool,nan2021entity,goodrich2019assessing,shuster2021retrieval}. However, in mobile task automation, actions of a mobile GUI agent lack a ground truth and its semantic ambiguity makes verification exceptionally difficult. \sys effectively handles this challenge by designing a domain-specific language (DSL, \autoref{sec:dsl}) and developer library (\autoref{sec:lib}) that effectively capture the underlying semantics of these ambiguous GUI actions.

\textbf{Leveraging Formal Logic for AI Safety.}
Recent research has explored the use of formal logic to ensure the safety and reliability of AI agents, particularly in areas like planning and control~\cite{jason2023grounding, ziyi2024plugin, yi2024selp, obi2025safeplan, nl2ltl, nl2spec}. Notably,
SELP~\cite{yi2024selp} ensures safe and efficient robot task planning by integrating constrained decoding and domain-specific fine-tuning.
Ziyi et al.~\cite{ziyi2024plugin} proposed a safety module for LFM-based robot agents by converting natural language safety constraints into Linear Temporal Logic. However,
while previous works typically focus on agents with a pre-defined safety concerns such as map navigation or indoor task planning ~\cite{jason2023grounding, ziyi2024plugin, yi2024selp, obi2025safeplan}, our system targets agents with an enormous number of environment, i.e., mobile apps.
To handle different environments of mobile apps, our optimized autoformalization algorithm effectively translates safety requirements for diverse set of mobile tasks.
\section{Discussion \& Future Work}
\label{sec:discussion}



\textbf{Piggybacking on App Development Frameworks.}
The current \sys{} implementation relies on developer-defined state and transition definition. However, given that modern mobile development frameworks like React Native, Jetpack Compose, and SwiftUI~\cite{react, jetpack, swift} already incorporate built-in state management, there is significant potential to "piggyback" on these existing mechanisms. Integrating \sys{} in this way could streamline the development process and reduce the overhead associated with adopting \sys{}.

\textbf{Automated State Definition via LFM.}
Another approach for simplifying \sys{}'s integration is automating the state definition process using LFMs. Existing GUI agents~\cite{autodroid, mobilegpt, appagent} often employ LFM-powered screen analysis to automatically annotate GUI elements and infer application state. A similar approach could be adapted to identify and define candidate state predicates for \sys{}. However, while this automation could reduce developer effort, it introduces inherent risks due to the probabilistic nature of LFMs. Relying on potentially inaccurate LFM-derived state definitions could compromise the system's reliability. Therefore, while LFM-based automation offers a potential convenience, the current approach of developer-defined states ensures the precision necessary for a robust verification system.

\textbf{Automated Predicate Memory Construction.}
As demonstrated in the evaluation, \sys{}-Warm's performance significantly benefits from the pre-populated predicate memory.  However, the current implementation relies on \sys{}-Cold to successfully complete tasks before they can be added to the memory, creating a performance bottleneck. One promising approach involves leveraging direct user feedback. By simply asking the user whether a task was completed successfully or not, we can obtain a ground-truth verification result. This user-provided ground truth can then be used to selectively populate the predicate memory. This approach would bypass the \sys{}-Cold bottleneck and accelerate the creation of a robust and accurate predicate memory, leading to improved overall performance.

\textbf{Interactive Feedback \& Dual Verification}
While \sys{} achieves high verification accuracy, it cannot guarantee 100\% safety. This is critical in high-risk scenarios like financial transactions where even a single error is unacceptable. To address these limitations, we propose human-in-the-loop strategies. Specifically, we recommend two complementary safeguards:  (1) interactive rule-level confirmation that prompts users for confirmation when rules are repeatedly violated, enabling quick human identification of potential False Positives. (2) Dual verification requiring human approval for critical operations (payments, data transfers) regardless of \sys's verification outcome---which guards against potential False Negatives. These mechanisms provide practical safeguards for commercial deployment while preserving \sys's automation benefits.






\section{Conclusion}
This paper introduced \textsf{VeriSafe Agent (VSA)}, a novel logic-based verification system designed to enhance the safety and reliability of Mobile GUI Agents. Through \sys, we have demonstrated that logic-driven verification can effectively safeguard GUI agents against the inherent uncertainties of LFM-based task automation. As AI-driven tasks 
become increasingly integrated into our daily life, \sys presents a crucial step toward safer use of AIs.
\begin{acks}
This work was partly supported by the Institute of Information \&
Communications Technology Planning \& Evaluation (IITP) grant (RS-2019-II190421, RS-2022-II221045, RS-2025-25442569, RS-2025-02305705, RS-2020-II201819, RS-2023-00232728), the National Research Foundation of Korea(NRF) grant (RS-2024-00338454, RS-2021-NR060080, RS-2025-00522352, RS-2024-00347516), and K-Startup grant (20144069) funded by Korea government (MSIT).
\end{acks}

\bibliographystyle{unsrtnat}
\bibliography{references}
\end{document}